\documentclass[conference, 10pt, twoside]{IEEEtran}
\IEEEoverridecommandlockouts
\normalsize

\usepackage{cite}
\usepackage{url}
\usepackage{hyperref}
\hypersetup{colorlinks=true, linkcolor=red, citecolor=red, urlcolor=blue} 
\usepackage{amssymb}
\usepackage{amsmath}
\usepackage{array}
\usepackage{amsthm}
\allowdisplaybreaks 
\usepackage{autobreak}
\usepackage{mathptmx}
\usepackage{mathtools, cuted}
\usepackage{algpseudocode}
\usepackage[ruled, vlined, linesnumbered]{algorithm2e}
\usepackage{graphicx}
\usepackage{bbm}
\usepackage{enumitem}
\setlist[itemize]{leftmargin=*}
\usepackage[table, dvipsnames]{xcolor}
\usepackage{subcaption}
\allowdisplaybreaks

\newtheorem{Theorem}{Theorem}

\newtheorem{Corollary}{Corollary}

\newtheorem{Remark}{Remark}

\newcommand{\rs}{\!\!}
\newcolumntype{C}[1]{>{\centering \arraybackslash}p{#1}}
\newcommand{\bblue}{\textcolor{black}}

\usepackage{setspace}
\usepackage{acronym}
\acrodef{ml}[ML]{machine learning}
\acrodef{fl}[FL]{federated learning}
\acrodef{hfl}[HFL]{hierarchical federated learning}
\acrodef{bs}[BS]{base station}
\acrodef{isp}[ISP]{(wireless) internet service provider}
\acrodef{ue}[UE]{user equipment}
\acrodef{es}[ES]{edge server}
\acrodef{csp}[CSP]{content service provider}
\acrodef{rawhfl}[RawHFL]{\underline{r}esource-\underline{aw}are \underline{h}ierarchical \underline{f}ederated \underline{l}earning}
\acrodef{hfedavg}[H-FedAvg]{hierarchical federated averaging}
\acrodef{sgd}[SGD]{stochastic gradient descent}
\acrodef{cpu}[CPU]{central processing unit}
\acrodef{prb}[pRB]{physical resource block}
\acrodef{snr}[SNR]{signal-to-noise-ratio}
\acrodef{lp}[LP]{linear programming}
\acrodef{fpp}[FPP]{floating point precision}
\acrodef{cdf}[CDF]{cumulative distribution function}
\acrodef{drl}[DRL]{deep reinforcement learning}

\title{Resource-Aware Hierarchical Federated Learning for Video Caching in Wireless Networks \vspace{-0.1in}}  
\author{
\IEEEauthorblockN{Md Ferdous Pervej and Andreas F. Molisch} \vspace{-0.15in}\\
\IEEEauthorblockA{University of Southern California, CA $90089$, USA; Email: \tt{\{pervej, molisch\}@usc.edu}}
\thanks{This work was supported by NSF-IITP Project $2152646$.}
\thanks{This work has been accepted for publication in the IEEE International Conference on Communications (ICC), Denver, CO, 2024. \copyright 2024 IEEE. Personal use of this material is permitted. Permission from IEEE must be obtained for all other uses, in any current or future media, including reprinting/republishing this material for advertising or promotional purposes, creating new collective works, for resale or redistribution to servers or lists, or reuse of any copyrighted component of this work in other works.}
\vspace{-0.2in}
}

\begin{document}
\maketitle
\IEEEpeerreviewmaketitle

\begin{abstract}
Video caching can significantly improve backhaul traffic congestion by locally storing the popular content that users frequently request. A privacy-preserving method is desirable to learn how users' demands change over time.
As such, this paper proposes a novel \underline{r}esource-\underline{aw}are \underline{h}ierarchical \underline{f}ederated \underline{l}earning (RawHFL) solution to predict users' future content requests under the realistic assumptions that content requests are sporadic and users' datasets can only be updated based on the requested content's information. 
Considering a partial client participation case, we first derive the upper bound of the global gradient norm that depends on the clients' local training rounds and the successful reception of their accumulated gradients over the wireless links.
Under delay, energy and radio resource constraints, we then optimize client selection and their local rounds and \ac{cpu} frequencies to minimize a weighted utility function that facilitates RawHFL's convergence in an energy-efficient way. 
Our simulation results show that the proposed solution significantly outperforms the considered baselines in terms of prediction accuracy and total energy expenditure.
\end{abstract}

\begin{IEEEkeywords}
Federated learning, resource-aware hierarchical federated learning, resource optimization, video caching.
\end{IEEEkeywords}

\section{Introduction}


\noindent
Video streaming is the dominant source of data traffic, and  $3$ out of $5$ video views occur on wireless devices \cite{loh2022youtube}.
As such, content caching \cite{golrezaei2013femtocaching} can become an integral part of modern wireless networks since it can save the backhaul transmission bandwidth and reduce network congestion by ignoring repetitive extractions of the same few popular videos the users repeatedly request from the far-away cloud.


Two major design components of an efficient video caching platform are content placement and content delivery \cite{liu2016caching}. 
Knowing which content the users will request in the near future can crucially help the provider in the content placement phase.
However, it is often challenging to predict content popularity as it changes rapidly.
Besides, many users may have their individual preferences for specific types of content that are not necessarily globally popular.
While \ac{ml} may accurately predict content popularity or user-specific content demand, the need for immense number of data samples for training the \ac{ml} model poses challenges in wireless video caching platforms.

A further challenge arises from requirements for privacy and/or protection of business secrets. 
The \ac{ue} makes a content request using its serving \ac{bs} to the \ac{csp} in wireless networks. 
On the one hand, the \ac{ue} and the \ac{csp} do not reveal the exact content ID/information to the \ac{bs}, the former because they want to protect their privacy, and the latter because \ac{csp} and the \ac{isp} operating the \ac{bs} are often competitors.
On the other hand, the spatial information of the \acp{ue} is only known to the \acp{isp}, which does not want to convey it to the \ac{csp}.
As such, privacy-preserving coordination among the \acp{ue}, \acp{isp} and \ac{csp} is required.
Therefore, distributed and privacy-preserving \ac{fl} \cite{mcmahan2017communication} is ideal for video caching in wireless networks.

Some existing literature \cite{qiao2022adaptive,wang2019edge,jiang2021federated,li2023community,khanal2022route} acknowledged the need for privacy protection and proposed \ac{fl}-based solutions for content caching.
In \cite{qiao2022adaptive}, Qiao \textit{et al.} proposed a \ac{fl} solution for content caching in resource-constrained wireless networks using two separate \ac{drl} agents to find a subset of clients and their local training rounds.
Wang \textit{et al.} also leveraged a similar strategy \cite{wang2019edge}, where users and edge servers used two separate \ac{drl} agents to learn computation offloading and content placement strategies, respectively.
Both \cite{qiao2022adaptive} and \cite{wang2019edge} considered federated aggregation of the \ac{drl} agents.
Jiang \textit{et al.} used an offline individual content preference learning, followed by an adaptive context space partitioning and \ac{fl}-based popularity prediction algorithm in \cite{jiang2021federated}.
Li \textit{et al.} developed an attention-weighted \ac{fl} algorithm for device-to-device wireless networks where they partition the \ac{ue}s into groups based on mobility and social behaviors in \cite{li2023community}.
Khanal \textit{et al.} proposed a self-attention-based \ac{fl} algorithm for content caching in vehicular networks where the moving self-driving cars, roadside units and macro \ac{bs} work collaboratively to train the model in \cite{khanal2022route}.

While the above studies recognized the privacy concern in content caching, the cooperation among the three entities, i.e., \ac{ue}, \ac{bs} and \ac{csp}, was not addressed.  
Besides, the above studies assumed that the client's\footnote{We use the terms \ac{ue} and client interchangeably throughout the paper.} training dataset is readily available.
In reality, a \ac{ue}'s content requests are sporadic, and its dataset only contains the requested content's information.
Moreover, since practical networks have hierarchical architectures where the client only communicates with its immediate upper tier, i.e., the BS \cite{wang2022demystifying}, a \ac{hfl} solution is needed.


Motivated by the above facts, we propose a novel \ac{rawhfl} algorithm for predicting clients' future content requests.
We consider that \ac{ue}s' requests arrive based on their own requirements, and their datasets can only be updated with the requested content's information.
Incorporating the well-known system and data heterogeneity, we derive the convergence analysis that reveals that the global gradient norm depends on the successful reception of the clients' trained accumulated gradients and their local training rounds.
As such, we jointly optimize client selection and clients' local training rounds and \ac{cpu} cycles to minimize a weighted utility function that facilitates \ac{rawhfl}'s convergence and minimizes energy expense under delay, energy and radio resource constraints.
Our extensive simulation results validate that the proposed solution outperforms existing baselines in terms of test accuracy and energy expense.

\begin{table}[!t]
\caption{Summary of important variables}
\centering
\fontsize{8}{8} \selectfont
\begin{tabular}{|C {1.3cm}|C{6.5cm}|}
\hline
\textbf{Parameter}  & \textbf{Definitions}  \\ \hline 
$u$, $\mathcal{U}$ & User $u$, all user set \\ \hline 
    $b$, $\mathcal{B}$ & base station $b$, all BS set \\ \hline 
    $l$, $\mathrm{L}$  & $l^{\mathrm{th}}$ SGD round, upper bound for local SGD round \\ \hline 
    $e$, $E$   & $e^{\mathrm{th}}$ edge round, total edge round  \\ \hline 
    $k$, $K$  & $k^{\mathrm{th}}$ global round, total global round \\ \hline 
    $t$ & $t^{th}$ discrete slot at which UE may request content \\\hline
    $\mathcal{U}_b$, $\bar{\mathcal{U}}_b^{k,e}$ & BS $b$'s UE set; selected UE/client set of BS $b$ during edge round $e$ of global round $k$ \\ \hline 
    $\mathrm{L}_u^{k,e}$ & UE/client $u$'s local SGD round during edge round $e$ of global round $k$ \\ \hline 
    $g$, $G$ & Genre $g$; total genres \\ \hline
    $c_g$, $\mathcal{C}_g$, $\mathcal{C}$ & $c^{\mathrm{th}}$ content of genre $g$; all content set in genre $g$; entire content catalog \\ \hline
    $\Bar{C}$, $C$ & Total content in genre $g$; total content in the catalog \\ \hline
    $z$, $Z$, $\mathcal{Z}$ & $z^{\mathrm{th}}$ pRB; total pRBs; pRB set \\ \hline 
    $\mathcal{D}_{u,\mathrm{r}}^0$ & UE $u$'s initial historical dataset \\ \hline
    $p_{u,\mathrm{ac}}$ & UE $u$'s probability of being active (making a content request) \\ \hline
    $\mathrm{1}_{u,c_g}^t$ & Binary indicator function that defines whether $u$ requests content $c_g$ during slot $t$ \\ \hline
    $p_{u,g}$ & UE $u$'s preference to genre $g$ \\ \hline
    $\Upsilon$ & Dirichlet distribution's concentration parameter for the genre preference \\ \hline
    $\mathcal{D}_{u,\mathrm{p}}^{t}$, $\mathrm{D}_{u,\mathrm{p}}^{t}$ & UE $u$'s processed dataset; total samples in UE $u$'s processed dataset \\ \hline   
    $f_u(\cdot)$; $f_b(\cdot)$, $f(\cdot)$ & UE $u$'s loss function; BS $b$'s loss function; global loss function \\ \hline
    $\mathbf{w}_u^{k,e,l}$, $\mathbf{w}_b^{k,e}$, $\mathbf{w}^{k}$ & UE/Client $u$'s local model during SGD round $l$ of edge round $e$ of global round $k$; BS $b$'s edge model during edge round $e$ of global round $k$; global model during round $k$ \\ \hline 
    $g(\mathbf{w}_u^{k,e,l})$ & UE $u$'s gradient during $l^{\mathrm{th}}$ local round of $e^{\mathrm{th}}$ edge round of $k^{\mathrm{th}}$ global round \\ \hline
    $\tilde{\mathrm{g}}_u^{k,e}$ & UE $u$'s accumulated gradients during $e^{\mathrm{th}}$ edge round of $k^{\mathrm{th}}$ global round \\ \hline
    $\eta$ & Learning rate \\ \hline
    $\alpha_u$, $\alpha_b$ & UE $u$'s trained model/accumulated gradients' weight; BS $b$'s model/accumulated gradients' weight \\ \hline
    $\mathrm{1}_{u, \mathrm{sl}}^{k,e}$; $\mathrm{p}_{u, \mathrm{sl}}^{k,e}$ & Binary indicator function to define whether $u$ is selected in edge round $e$ of global round $k$; success probability of $\mathrm{1}_{u,\mathrm{sl}}^{k,e}$ \\ \hline
    $\mathrm{t}_{u,\mathrm{cp}}^{k,e}$; $\mathrm{e}_{u,\mathrm{cp}}^{k,e}$ & UE $u$'s local model training time and energy overheads during edge round $e$ of global round $k$ \\ \hline
    $\mathrm{t}_{u,\mathrm{up}}^{k,e}$; $\mathrm{e}_{u,\mathrm{up}}^{k,e}$ & UE $u$'s accumulated gradient offloading time and energy overheads during edge round $e$ of global round $k$ \\ \hline
    $\mathrm{t_{th}}$ & Deadline threshold to finish one edge round \\ \hline
    $P_{u,\mathrm{tx}}$ & Transmission power of $u$ \\ \hline 
    $\mathrm{e}_{u,\mathrm{bd}}$ & Energy budget of $u$ for each edge round \\ \hline
    $\mathrm{1}_{u,\mathrm{sc}}^{k,e}$; $\mathrm{p}_{u,\mathrm{sc}}^{k,e}$ & Binary indicator function to define whether accumulated gradient of $u$ is received by the \ac{bs}; success probability of $\mathrm{1}_{u,\mathrm{sc}}^{k,e}$ \\ \hline
    $\phi$; $s$ & Floating point precision; client's uplink payload size for the accumulated gradients \\ \hline
    $\varphi_b^{k,e}$ & Utility function for the joint optimization problem \\ \hline
\end{tabular}
\label{tableofVariable}
\end{table}

\section{System Model}

\subsection{Network Model 
}
\noindent
We consider a cache-enabled wireless network, consisting of distributed \acp{ue}, \acp{bs} and a \ac{csp}. 
Denote the UE set $\mathcal{U}=\{u\}_{u=0}^{U-1}$ and the BS set by $\mathcal{B}=\{b\}_{b=0}^{B-1}$.
Besides, denote the UEs that are associated to BS $b$ by $\mathcal{U}_b$ such that $\mathcal{U} = \bigcup_{b=0}^{B-1} \mathcal{U}_b$.
The system operates in discrete time slot $t=1, \dots, T$, where the duration between two slots is $\kappa$ seconds.
The CSP has a fixed content catalog, denoted by $\mathcal{C}=\{\mathcal{C}_g\}_{g=0}^{G-1}$, where $\mathcal{C}_g = \{c_g\}_{c=0}^{\bar{C}-1}$ is the content set of genre $g$.
Besides, denote the total number of content in the catalog by $C = \Bar{C} G$.
Each \ac{bs} is equipped with an \ac{es} 
that has a limited cache storage and computation capability, and is under the control of the \ac{csp}.
Consequently, the \ac{bs} has no information about what content is stored in the \ac{es}.
To keep its operational information private, the \ac{csp} assigns temporal tag IDs to the original content ID that it shares with the \ac{bs}. 
The mappings between the actual content IDs and the tagged IDs are only known to the \ac{csp} and the \acp{ue}.
Moreover, the \ac{csp} can periodically change these mappings to prevent the BS from learning the actual content information.
Note that since each \ac{bs} is equipped with a distinct \ac{es}, we use the same notation $b$ to represent the \ac{es} of the $b^{\mathrm{th}}$ \ac{bs} for brevity.

The network has $\bar{\mathrm{Z}}$ Hz bandwidth allocated for performing the \ac{fl} task,
which is further divided into $\mathcal{Z}=\{z\}_{z=0}^{Z-1}$ orthogonal \acp{prb}.
The frequency reuse factor is $1$, i.e., each BS utilizes the same pRB set.
Besides, we assume that the \acp{bs} collaborate to find node associations and pRB allocations such that inter-cell interference is fully mitigated.
A list of the important notations used in this paper is summarized in Table \ref{tableofVariable}.

\subsection{Content Request Model 
}

\noindent 
This work assumes that each UE has a small initial historical raw dataset, denoted by $\mathcal{D}_{u,\mathrm{r}}^0$.
During slot $t$, a \ac{ue} may request a content from the \ac{csp} with probability $p_{u,\mathrm{ac}}$, which can be chosen according to the \ac{ue}'s \textit{activity level}.
Given that the \ac{ue} is \textit{active}, we use the binary indicator function $\mathrm{1}_{u,c_g}^{t} \in \{0,1\}$ to denote \textit{which} content the UE requests in that slot. 
The \ac{ue} stores the requested content's information in its local raw dataset that evolves as follows  
\begin{equation}
    \label{datasetEvol}
    \begin{aligned}
        \mathcal{D}_{u,\bblue{\mathrm{r}}}^t &\coloneqq 
        \begin{cases}
            \mathcal{D}_{u,\bblue{\mathrm{r}}}^{t-1} \bigcup \big\{ \mathbf{x}(\mathrm{1}_{u,c_g}^t), \mathbf{y}(\mathrm{1}_{u,c_g}^t) \big\},   & \text{if $u$ is active}, \\
            \mathcal{D}_{u,\bblue{\mathrm{r}}}^{t-1}, & \text{otherwise},
        \end{cases}\rs,
    \end{aligned}
\end{equation}
where $\mathbf{x}(\mathrm{1}_{u,c_g}^t)$ and $\mathbf{y}(\mathrm{1}_{u,c_g}^t)$ are the feature and label vectors, respectively, for the requested content.
As such, our dataset acquisition mimics the natural data sensing in real-world applications, where the dataset sizes are time-varying  \cite{pervej2022mobility,hosseinalipour2023parallel}.

Besides, each \ac{ue} follows a popularity-preference tradeoff in its content request model.
More specifically, the \acp{ue} have their own independent genre preferences. 
Denote \ac{ue} $u$'s preference for genre $g$ by $p_{u,g}$ such that $0\leq p_{u,g} \leq 1$ and  $\sum_{g=0}^{G-1} p_{u,g}=1$.
We model the genre preference using symmetric Dirichlet distribution $\mathrm{Dir}(\pmb{\Upsilon})$, where the $\pmb{\Upsilon}$ is the concentration parameter that controls the skewness \cite{pervej2023Resource}. 
Initially, the \ac{ue} requests the most popular content from the selected genre $g$. 
For the subsequent request, it requests the most similar content\footnote{We consider each content's distinctive feature set and calculate the cosine similarity of the content within the same genre.} to the previously requested content in the same genre $g$ with probability $\upsilon_u$ and the most popular content of a different genre $g' \neq g$ with probability $(1-\upsilon_u)$.

\subsection{Hierarchical Federated Learning: Preliminaries 
}
\noindent
The central server wants to train an \ac{ml} model, parameterized by $\mathbf{w} \in \mathbb{R}^d$, to predict the future content requests of the \ac{ue}s.
While the \acp{ue} will not reveal their content preferences, they are eager to participate in the model training and share their predictions with the \ac{es}.
Without loss of generality, during slot $t$, each \ac{ue} processes its local raw dataset and prepares a processed dataset $\mathcal{D}_{u,\bblue{\mathrm{p}}}^t = \{\mathbf{x}_{u}^a, \mathbf{y}_{u}^a\}_{a=0}^{\mathrm{D}_{u,\bblue{\mathrm{p}}}^t - 1}$, where $(\mathbf{x}_{u}^a, \mathbf{y}_u^a)$ is the $a^{\mathrm{th}}$ (processed) training samples and $\mathrm{D}_{u,\bblue{\mathrm{p}}}^t$ is the total training samples.
Using their processed datasets, each \ac{ue} wants to minimize the following loss function.
\begin{equation}
\label{localLossFunc}
\begin{aligned}
    f_u (\mathbf{w} | \mathcal{D}_{u,\bblue{\mathrm{p}}}^t) \coloneqq [1/\bblue{\mathrm{D}_{u,\mathrm{p}}^t}] \sum\nolimits_{(\mathbf{x}_{u}^a, \mathbf{y}_u^a) \in \mathcal{D}_{u,\bblue{\mathrm{p}}}^t} \mathrm{l}\left(\mathbf{w} | (\mathbf{x}_{u}^a, \mathbf{y}_u^a) \right),
\end{aligned}
\end{equation}
where $\mathrm{l}\left(\mathbf{w} | (\mathbf{x}_{u}^a, \mathbf{y}_u^a)\right)$ is the loss associated with the $a^{\mathrm{th}}$ data sample.

In \ac{hfl} \cite{liu2020client}, the immediate upper tier of the clients, i.e., the \ac{es}, wishes to minimize 
\begin{equation}
\label{hflGlobalLoss}
    \begin{aligned}
        f_{b} (\mathbf{w} | \mathcal{D}_{b}^t) \coloneqq \sum\nolimits_{u \in \mathcal{U}_b} \alpha_u f_u (\mathbf{w} | \mathcal{D}_{u}^t),
    \end{aligned}
\end{equation}
where $\alpha_u$ is the weight of the $u^{\mathrm{th}}$ client in BS $b$ and $\mathcal{D}_b^t \coloneqq \bigcup_{u \in \mathcal{U}_b} \mathcal{D}_{u,\bblue{\mathrm{p}}}^t$.
Besides, the upper tier of the \acp{es}, i.e., the central server, aims at minimizing the following global loss function.
\begin{equation}
\begin{aligned}
    f(\mathbf{w} | \mathcal{D}^t) \coloneqq \sum\nolimits_{b=0}^{B-1} \alpha_b f_b (\mathbf{w} | \mathcal{D}_{b}^t) = \sum\nolimits_{b=0}^{B-1} \alpha_b \sum\nolimits_{u \in \mathcal{U}_b} \alpha_u f_u (\mathbf{w} | \mathcal{D}_{u}^t),
\end{aligned}
\end{equation}
where $\alpha_b$ is the weight of the \ac{es} of the $b^{\mathrm{th}}$ \ac{bs} at the central server and $\mathcal{D}^t \coloneqq \bigcup_{b=0}^{B-1} \mathcal{D}_b^t$.
Moreover, due to the dynamic changes in the local datasets, the optimal global model $\mathbf{w}^{*}$ is not necessarily stationary \cite{hosseinalipour2023parallel}.
As such, the central server seeks a sequence of $\mathbf{w}^{*}$'s, where 
\begin{equation}
\begin{aligned}
    \mathbf{w}^{*} &= \underset{\mathbf{w} }{\text{arg min}} \quad f(\mathbf{w} | \mathcal{D}^t), \quad \forall t.
\end{aligned}
\end{equation}

\section{Resource-Aware Hierarchical Federated Learning: Algorithm and Convergence}

\subsection{Resource-Aware Hierarchical Federated Learning Model 
}

\noindent
Similar to general \ac{hfl} \cite{liu2020client,pervej2023hierarchical,wang2022demystifying}, in our proposed \ac{rawhfl}, the clients, the 
\ac{es} and the central server perform local, edge and global rounds.
The nodes in each tier perform their local training before sending their updated models to their respective upper levels.
Besides, due to resource constraints, we consider that each \ac{es} selects only a subset of the clients to participate in model training.
Let $\bar{\mathcal{U}}_b^{k,e} \subseteq \mathcal{U}_b$ denote the selected client set of the \ac{es} of \ac{bs} $b$ during the $e^{\mathrm{th}}$ edge round of the $k^{\mathrm{th}}$ global round.
Denote the client's participation by
\begin{equation}
    \begin{aligned}
        \mathrm{1}_{u,\mathrm{sl}}^{k,e} &= 
        \begin{cases}
            1, &\text{with probability } \mathrm{p}_{u, \mathrm{sl}}^{k,e},\\
            0, &\text{otherwise},
        \end{cases}.
    \end{aligned}
\end{equation}

In each edge round, the \ac{es} sends its model to its \ac{bs}. 
The \ac{bs} then broadcasts\footnote{Similar to existing studies \cite{hosseinalipour2023parallel,pervej2023Resource}, we ignore the downlink transmission time as the \ac{bs} can use higher transmission power and entire bandwidth.} its available model to all $u \in \bar{\mathcal{U}}_b^{k,e}$, who synchronize their local models as
\begin{equation}
    \mathbf{w}_u^{k,e,0} \gets \mathbf{w}_b^{k,e}.
\end{equation}
The client takes a \ac{sgd} step to minimize (\ref{localLossFunc}) and updates its model as
\begin{equation}
    \label{localModUp}
    \begin{aligned}
        \mathbf{w}_{u}^{k,e,l+1} 
        &= \mathbf{w}_{u}^{k, e, l} - \eta g(\mathbf{w}_{u}^{k, e, l}),
    \end{aligned}
\end{equation}
where $\eta$ is the learning rate. 
Particularly, each client has $\mathrm{t_{th}}$ seconds and $\mathrm{e}_{u,\mathrm{bd}}$ Joules of time and energy budgets to spend in each edge round $e$.
As such, each client performs $\mathrm{L}_{u}^{k,e}$, where $1 \leq \mathrm{L}_{u}^{k,e} \leq \mathrm{L}$, \ac{sgd} rounds, which can be different for different \acp{ue}.
Therefore, we calculate the local computation time as
\begin{equation}
    \mathrm{t}_{u,\mathrm{cp}}^{k,e} = \mathrm{L}_{u}^{k,e} \times \mathrm{n} \bar{\mathrm{n}} \mathrm{c}_u \mathrm{D}_u / f_u^{k,e},
\end{equation}
where $\mathrm{n}$, $\bar{\mathrm{n}}$, $\mathrm{c}_u$, $\mathrm{D}_u$ and $f_u^{k,e}$ are the number of mini-batches, batch size, number of \ac{cpu} cycle to compute $1$-bit data, data sample size in bits and the \ac{cpu} frequency.
Besides, the energy expense for performing these $\mathrm{L}_{u}^{k,e}$ \ac{sgd} rounds is \cite{qiao2022adaptive, pervej2023hierarchical,hosseinalipour2023parallel}
\begin{equation}
\begin{aligned}
    \mathrm{e}_{u,\mathrm{cp}}^{k,e} 
    &= \mathrm{L}_{u}^{k,e} \times 0.5\zeta \mathrm{n} \bar{\mathrm{n}} \mathrm{c}_u \mathrm{D}_u (f_u^{k,e})^2,
\end{aligned}
\end{equation}
where $0.5\zeta$ is the effective capacitance of the CPU chip.


After finishing local training, each client offloads its accumulated gradient $\tilde{\mathrm{g}}_u^{k,e} \coloneqq \sum_{l=0}^{\mathrm{L}_u^{k,e}-1} g(\mathbf{w}_u^{k,e,l})$ to the \ac{bs}, which then forwards it to its \ac{es}.
This accumulated gradient incurs a wireless payload size of $\mathrm{s} = d \times (\phi+1)$ bits \cite{pervej2023Resource}, where $\phi$ is the \ac{fpp}.
The required time to offload $\tilde{\mathrm{g}}_u^{k,e}$ is 
\begin{equation}
    \mathrm{t}_{u,\mathrm{up}}^{k,e} = \mathrm{s}/ \big(\omega \log_2 \big[1 + \gamma_{u}^{k,e}\big] \big),
\end{equation}
where $\omega$ is the \ac{prb} size and $\gamma_u^{k,e}$ is the \ac{snr}, which is calculated as\footnote{Since practical networks now offer enough diversity against small-scale fading, we dropped the small-scale fading channel factor $|h_u^{k,e}|^2$.} 
\begin{equation}
\label{snr}
    \gamma_{u}^{k,e} = \beta_{u}^{k,e} \zeta_u^{k,e} \bblue{P_{u,\mathrm{tx}}} / (\omega \varsigma^2), 
\end{equation}
where $\bblue{P_{u,\mathrm{tx}}}$ is the uplink transmission power of the $u^{\mathrm{th}}$ client. 
Besides, $\beta_u^{k,e}$ and $\zeta_u^{k,e}$ are the path loss and log-Normal shadowing. 
Furthermore, $\varsigma^2$ is the variance of the circularly symmetric zero-mean Gaussian distributed random noise. 
Moreover, the required energy expense to offload $\tilde{\mathrm{g}}_u^{k,e}$ is calculated as
\begin{equation}
    \mathrm{e}_{u,\mathrm{up}}^{k,e} = \mathrm{s} \bblue{P_{u,\mathrm{tx}}} / \big(\omega \log_2 \big[1 + \gamma_{u}^{k,e}\big]\big).
\end{equation}

\bblue{Note that due to the subset client selection, each \ac{es} minimizes $f_{b} (\mathbf{w} | \cup_{u \in \bar{\mathcal{U}}_b^{k,e}} \mathcal{D}_{u, \mathrm{proc}}^t ) \coloneqq \sum\nolimits_{u \in \bar{\mathcal{U}}_b^{k,e}} \alpha_u f_u (\mathbf{w} | \mathcal{D}_{u,\mathrm{proc}}^t)$, where $\alpha_u \coloneqq 1/|\bar{\mathcal{U}}_b^{k,e}|$.}
During the edge aggregation time, each \ac{es} updates its edge model using the $\tilde{\mathrm{g}}_u^{k,e}$'s from all of its selected clients as
\begin{equation}
\label{edgeUpdateRule}
\begin{aligned}
    \mathbf{w}_b^{k, e+1} 
    &= \mathbf{w}_{b}^{k,e} - \eta \sum\nolimits_{u \in \bar{\mathcal{U}}_b^{k,e}} \alpha_u \big[\mathrm{1}_{u,\mathrm{sc}}^{k,e}/\mathrm{p}_{u,\mathrm{sc}}^{k,e}\big]
    \tilde{\mathrm{g}}_u^{k, e},
\end{aligned}
\end{equation}
where $\mathrm{1}_{u,\mathrm{sc}}^{k,e}$ is a binary indicator that defines whether $\tilde{\mathrm{g}}_u^{k,e}$ is received at the \ac{bs} during the aggregation time and is defined as
\begin{equation}
    \begin{aligned}
        \mathrm{1}_{u,\mathrm{sc}}^{k,e} &= 
        \begin{cases}
            1, &\text{with probability } \mathrm{p}_{u, \mathrm{sc}}^{k,e},\\
            0, &\text{otherwise},
        \end{cases}.
    \end{aligned}
\end{equation}
Each \ac{bs} then broadcasts their respective \ac{es}' updated model to their selected clients, and the clients perform local training and offload back the accumulated gradients. 
This process repeats for $E$ edge rounds, after which the \acp{es} send their updated models to the central server that aggregates the received edge models as
\begin{equation}
\label{globalUpdateRule}
\begin{aligned}
    \rs\rs \rs \rs \mathbf{w}^{k+1} 
    \rs= \mathbf{w}^k \rs - \eta \sum\nolimits_{e=0}^{E-1} \sum\nolimits_{b=0}^{B-1} \rs \alpha_b \sum\nolimits_{u \in \bar{\mathcal{U}}_b^{k,e}} \alpha_u \big[\mathrm{1}_{u,\mathrm{sc}}^{k,e} / \mathrm{p}_{u,\mathrm{sc}}^{k,e}\big] \tilde{\mathrm{g}}_u^{k, e} \rs. \rs\rs\rs\rs
\end{aligned}   
\end{equation}
\bblue{Note that the global loss of \ac{rawhfl} is $f(\mathbf{w}^k | \cup_{b=0}^{B-1} \cup_{u \in \bar{\mathcal{U}}_b^{k,e}} \mathcal{D}_{u,\mathrm{proc}}^t) \coloneqq \sum\nolimits_{b=0}^{B-1} \alpha_b f_{b} \big(\mathbf{w}^k | \cup_{u \in \bar{\mathcal{U}}_b^{k,e}} \mathcal{D}_{u, \mathrm{proc}}^t \big)$, which may differ from (\ref{hflGlobalLoss}) if $\mathcal{U}_b^{k,e} \subset \mathcal{U}_b$.} 
The central server then sends this updated model to the \acp{es}, who perform their edge rounds following the above process. 
Algorithm \ref{rawHFLAlg} summarizes these steps.
\begin{algorithm}[!t]
\small
\SetAlgoLined 
\DontPrintSemicolon
\KwIn{Global model: $\mathbf{w}^0$; total global round $K$, number of edge rounds $E$ }
\For{$k=0$ to $K-1$}{
    \For {$b=1$ to $B$ in parallel}{
        Receives updated global model $\mathbf{w}_b^{k,0} \gets \mathbf{w}^k$ \;
        \For {$e=0$ to $E-1$} {
            BS $b$ receives optimized $\mathrm{1}_{u, \mathrm{sl}}^{k,e}, \mathrm{L}_{u}^{k,e}$ and $f_{u}^{k,e}$ \tcp*{\textit{c.f.} (\ref{originalOptimProb})}
            \For{$u$ in $\bar{\mathcal{U}}_b^{k,e}$ in parallel}{
                Get updated edge model $\mathbf{w}_{u}^{k,e,0} \gets \mathbf{w}_b^{k,e}$ \;
                Perform $\mathrm{L}_u^{k,e}$ mini-batch SGD rounds and get updated local model based on (\ref{localModUp}) \;
                Offload accumulated gradients $\tilde{\mathrm{g}}_u^{k,e}$ to the \ac{bs}\;
            }
            Update edge model $\mathbf{w}_b^{k, e+1}$ using update rule in (\ref{edgeUpdateRule})
        }
    }
    Update global model $\mathbf{w}^{k+1}$ based on the update rule in (\ref{globalUpdateRule}) \;
}
\KwOut{Trained global model $\mathbf{w}^K$}
\caption{RawHFL Algorithm}
\label{rawHFLAlg}
\end{algorithm}

\subsection{Convergence of RawHFL 
}

\noindent
We make the following standard assumptions \cite{liu2020client, pervej2023hierarchical, wang2022demystifying}
\begin{enumerate}
    \item The loss functions are $\beta$-smooth.
    \item The mini-batch gradients are unbiased. The variance of the gradients is bounded, i.e., $\Vert g(\mathbf{w}) - \nabla f_u(\mathbf{w}) \Vert^2 \leq \sigma^2$.
    \item The stochastic gradients in different local epochs, client selection and accumulated gradient offloading in edge rounds are independent.  
    \item The divergence between the two interconnected tiers' loss functions is bounded. For all $u$, $b$ and $\mathbf{w}$, 
    \begin{align*}
        &\sum\nolimits_{u \in \mathcal{U}_b} \alpha_u \Vert \nabla f_u(\mathbf{w}) - \nabla f_b(\mathbf{w}) \Vert^2 \leq \epsilon_0^2,\\
        &\rs\rs\rs\rs\rs\rs \bblue{\sum\nolimits_{b=0}^{B-1} \alpha_b \Big\Vert \rs \sum_{u \in \bar{\mathcal{U}}_b^{k,e}} \rs \rs \rs \alpha_u \nabla \tilde{f}_{u} (\mathbf{w}) - \sum\nolimits_{b'=0}^{B-1} \alpha_{b'} \rs \rs \rs \rs \sum_{u' \in \bar{\mathcal{U}}_{b'}^{k,e}} \rs\rs\rs \alpha_{u'} \nabla \tilde{f}_{u'} (\mathbf{w}) \Big\Vert^2 \leq \epsilon_1^2,} 
    \end{align*}
    \bblue{where $\nabla \tilde{f}_u (\mathbf{w}) \coloneqq \sum_{l=0}^{\mathrm{L}_u^{k,e} - 1} \nabla f_u (\mathbf{w})$.}
\end{enumerate}

\begin{Theorem}
\label{theorem1}
Suppose the above assumptions hold. 
When \bblue{$\eta < \mathrm{min}\left\{\frac{1}{2\sqrt{5} \beta \mathrm{L}}, \frac{1}{\beta E \mathrm{L}} \right\}$}, the average global gradient norm is upper-bounded as
\begin{align}
\label{theorem1_eqn}
    & \frac{1}{K} \sum_{k=0}^{K-1} \mathbb{E} \big[\Vert \nabla f(\mathbf{w}^k) \Vert^2 \big]
    \leq \frac{2}{\eta K} \sum_{k=0}^{K-1} \frac{1}{\Omega^k} \Big\{ \mathbb{E} [ f(\mathbf{w}^k) ] - \mathbb{E} [ f(\mathbf{w}^{k+1}) ] \Big\} \nonumber\\
    & + \frac{2 \beta \eta \mathrm{L} \sigma^2}{K} \sum_{k=0}^{K-1} \frac{\mathrm{N}_1^k}{\Omega^k} + \frac{18 E \beta^2 \epsilon_0^2 \eta^2 \mathrm{L}^3}{K} \sum_{k=0}^{K-1} \frac{\mathrm{N}_2}{\Omega^k} + \nonumber\\
    &\frac{\bblue{20 \mathrm{L}} \beta^2 \epsilon_1^2 \eta^2 E^3}{K} \sum_{k=0}^{K-1} \frac{1}{\Omega^k} + \frac{2 \beta \eta \mathrm{L}}{K} \sum_{k=0}^{K-1} \frac{1}{\Omega^k} \sum_{e=0}^{E-1} \sum_{b=0}^{B-1} \alpha_b \times \nonumber\\
    & \rs \rs \sum\nolimits_{u \in \bar{\mathcal{U}}_b^{k,e}} \alpha_u \mathrm{N}_u  \big[(1/\mathrm{p}_{u,\mathrm{sc}}^{k,e}) - 1 \big] \sum\nolimits_{l=0}^{\mathrm{L}_u^{k,e} - 1} \mathbb{E} \big[\big\Vert g_u (\mathbf{w}_u^{k, e, l}) \big\Vert^2 \big],
\end{align}
where \bblue{the expectations depend on clients' randomly selected mini-batches and $\mathrm{1}_{u,\mathrm{sc}}^{k,e}$'s. 
Besides,} $\Omega^k \coloneqq \sum_{e=0}^{E-1} \sum_{b=0}^{B-1} \alpha_b \sum_{u \in \bar{\mathcal{U}}_b^{k,e}} \alpha_u \mathrm{L}_{u}^{k,e}$, \bblue{$\mathrm{N}_1^k \coloneqq  60 \beta^3 \eta^3 E^3 \mathrm{L}^3 + 3 \beta \eta E \mathrm{L} +  \sum_{e=0}^{E-1} \sum_{b=0}^{B-1} \alpha_b  \left(\alpha_b + 4 E \mathrm{L} \beta \eta \right) \sum_{u \in \bar{\mathcal{U}}_b^{k,e}} \left(\alpha_u\right)^2$, $\mathrm{N}_2 \coloneqq 1 + 20 \beta^2 \eta^2 E^2 \mathrm{L}^2$ and $\mathrm{N}_{u} \coloneqq  E + 3 \beta \eta \mathrm{L} + 4 \beta \eta E \left(\alpha_u + 15 E \beta^2 \eta^2 \mathrm{L}^3 \right)$.}
\end{Theorem}

\begin{proof}[\textbf{Sketch of Proof}]
We start with the aggregation rule (\ref{globalUpdateRule}) and the $\beta$-smoothness assumption, \bblue{i.e., $ f(\mathbf{w}^{k+1}) \leq f(\mathbf{w}^k) - \eta \big<\nabla f(\mathbf{w}^k),  \sum_{e=0}^{E-1} \sum_{b=0}^{B-1} \alpha_b\sum_{u \in \bar{\mathcal{U}}_b^{k,e}} \alpha_u \frac{\mathrm{1}_{u,\mathrm{sc}}^{k,e}}{\mathrm{p}_{u,\mathrm{sc}}^{k,e}} \sum_{l=0}^{\mathrm{L}_{u}^{k,e}-1} g \big(\mathbf{w}_u^{k, e,l}\big) \big> + \frac{\beta \eta^2}{2} \big\Vert \sum_{e=0}^{E-1} \sum_{b=0}^{B-1} \alpha_b\sum_{u \in \bar{\mathcal{U}}_b^{k,e}} \alpha_u \frac{\mathrm{1}_{u,\mathrm{sc}}^{k,e}}{\mathrm{p}_{u,\mathrm{sc}}^{k,e}} \tilde{\mathrm{g}}_u^{k, e} \big\Vert^2.$}   
Then, we derive the upper bounds of the inner-product and $L_2$ norm terms\bblue{, which gives the following when $\eta \leq \frac{1}{\beta E \mathrm{L}}$.}
\begin{align}
\label{theorem1_Mid_Bound}
    &\rs\rs\rs \frac{1}{K} \rs \sum_{k=0}^{K-1} \rs \mathbb{E} \big[\big\Vert \nabla f(\mathbf{w}^k)\big\Vert^2 \big]
    \leq \frac{2}{\eta K} \sum_{k=0}^{K-1} \bigg[ \frac{\mathbb{E} [f(\mathbf{w}^k)] - \mathbb{E} \left[ f(\mathbf{w}^{k+1}) \right] }{\Omega^k} \bigg] + \nonumber\\
    &\rs\rs \frac{2 \beta \eta \sigma^2}{K} \sum_{k=0}^{K-1} \rs \Bigg[\frac{\sum_{e=0}^{E-1} \sum_{b=0}^{B-1} \left(\alpha_b\right)^2 \sum_{u \in \bar{\mathcal{U}}_b^{k,e}} \left(\alpha_u \right)^2 \mathrm{L}_u^{k,e} }{\Omega^k} \Bigg] \rs + \frac{2 \beta \eta E}{K} \times \nonumber\\
    &\rs\rs \sum_{k=0}^K \frac{1}{\Omega^k} \rs \sum_{e=0}^{E-1} \sum_{b=0}^{B-1} \rs \alpha_b \rs \rs\rs \sum_{u \in \bar{\mathcal{U}}_b^{k,e}} \rs\rs\rs\rs \alpha_u \mathrm{L}_u^{k,e} \sum_{l=0}^{\mathrm{L}_u^{k,e} - 1} \rs \bigg[\frac{1} {\mathrm{p}_{u,\mathrm{sc}}^{k,e}} - 1 \bigg] \mathbb{E} \big[\big\Vert g_u (\mathbf{w}_u^{k, e, l}) \big\Vert^2 \big] + \nonumber\\
    & \frac{2 \mathrm{L} \beta^2}{K} \sum_{k=0}^{K-1} \frac{1}{\Omega^k} \sum_{e=0}^{E-1} \sum_{b=0}^{B-1} \alpha_b \mathbb{E} \left[ \left\Vert \mathbf{w}^k - \mathbf{w}_{b}^{k,e} \right\Vert^2 \right] + \nonumber\\
    &\frac{2 \beta^2}{K} \sum_{k=0}^{K-1} \frac{1}{\Omega^k} \sum_{e=0}^{E-1} \sum_{b=0}^{B-1} \alpha_b \rs \rs \sum_{u \in \bar{\mathcal{U}}_b^{k,e}} \rs \rs \alpha_u \rs \bblue{\sum_{l=0}^{\mathrm{L}_u^{k,e} - 1}} \mathbb{E} \left[ \left\Vert \mathbf{w}_{b}^{k,e} - \mathbf{w}_{u}^{k,e,l} \right\Vert^2 \right].
\end{align}
\bblue{To that end, assuming \bblue{$\eta \leq \mathrm{min} \left\{\frac{1}{\beta E \mathrm{L}}, \frac{1}{3\sqrt{2} \beta \mathrm{L}} \right\}$}, we derive the upper bounds of the last term of (\ref{theorem1_Mid_Bound}), and similarly for the second last term using \bblue{$\eta < \mathrm{min}\left\{\frac{1}{2\sqrt{5} \beta \mathrm{L}}, \frac{1}{\beta E \mathrm{L}} \right\}$}.
Finally, we plug those terms in (\ref{theorem1_Mid_Bound}) and do some algebraic manipulations to reach (\ref{theorem1_eqn}).}
\end{proof}
\begin{Remark}
\label{remTheorem}
    In (\ref{theorem1_eqn}), the first term captures the changes in the global loss function in consecutive rounds. 
    The following terms with $\sigma^2$ come from the variance of the gradients.
    Besides, the following terms with $\epsilon_0^2$ and $\epsilon_1^2$ stem from the bounded divergence assumptions of the loss functions. 
    Finally, the last term appears from the wireless links between the \acp{ue} and the \acp{bs}.
    Moreover, when the accumulated gradients are received successfully, i.e., all $\mathrm{p}_{u, \mathrm{sc}}^{k,e}=1$, the last term becomes $0$.
\end{Remark}

\begin{Corollary}
\label{corol_1}
When $\mathrm{L}_u^{k,e}=\mathrm{L}$, $\forall u, k$ and $e$, the bound in Theorem \ref{theorem1} boils down to 
\begin{align*}
\label{corr0}
    & \frac{1}{K} \sum_{k=0}^{K-1} \mathbb{E} \big[\Vert \nabla f(\mathbf{w}^k) \Vert^2 \big]
    \leq \frac{2 ( \mathbb{E} [f(\mathbf{w}^0)] - \mathbb{E} [ f(\mathbf{w}^{K}) ])} {\eta E K \mathrm{L}} + \nonumber\\
    &[2 \beta \eta \sigma^2 / (E K) ] \sum\nolimits_{k=0}^{K-1} \mathrm{N}_1^k + 18 \beta^2 \epsilon_0^2 \eta^2 \mathrm{L}^2 \mathrm{N}_2 + \bblue{20} \beta^2 \epsilon_1^2 \eta^2 E^2 +\nonumber\\
    & \frac{2 \beta \eta}{E K} \sum_{k=0}^K \sum_{e=0}^{E-1} \sum_{b=0}^{B-1} \rs \alpha_b \rs \rs \rs \sum_{u \in \bar{\mathcal{U}}_b^{k,e}} \rs\rs\rs \alpha_u \mathrm{N}_u \rs\rs \sum_{l=0}^{\mathrm{L}_u^{k,e} - 1} \rs \bigg[\frac{1}{\mathrm{p}_{u,\mathrm{sc}}^{k,e}} - 1 \bigg] \mathbb{E} \big[\big\Vert g_u (\mathbf{w}_u^{k, e, l}) \big\Vert^2 \big].
\end{align*}
\end{Corollary}

\section{RawHFL: Joint Problem Formulation and Solutions}

\noindent
Theorem \ref{theorem1}, Corollary \ref{corol_1} and Remark \ref{remTheorem} show that the controllable terms in the convergence bound are the $\Omega^k$'s and the $\mathrm{p}_{u, \mathrm{sc}}^{k,e}$'s.
Besides, $\mathrm{p}_{u, \mathrm{sc}}^{k,e}$'s are intertwined with the \ac{fl} parameters ($\mathrm{L}_u^{k,e}$'s and $f_u^{k,e}$'s) and wireless factors that influence (\ref{snr}).
Furthermore, we assume that the accumulated gradients $\tilde{\mathrm{g}}_u^{k,e}$ is transmitted as a single wireless packet. 
The \ac{bs} can successfully decode the gradients without errors\footnote{This assumption is reasonable as wireless networks use hybrid automatic repeat request and error correction mechanisms \cite[Chap. $13$]{molisch2023wireless}.} if it receives the packet within the deadline $\mathrm{t_{th}}$.
As such, we denote $\mathrm{p}_{u,\mathrm{sc}}^{k,e} = \mathrm{Pr}\{[ \mathrm{t}_{u,\mathrm{cp}}^{k,e} + \mathrm{t}_{u,\mathrm{up}}^{k,e} \leq \mathrm{t_{th}}\}$.
The above facts inspire us to solve the following optimization problem to jointly select the clients and find their \ac{cpu} frequencies and local iterations at the beginning of each edge round $e$.
\begin{subequations}
\label{originalOptimProb}
\begin{align}
    &\tag{\ref{originalOptimProb}}
    \underset{\pmb{\mathrm{1}}_{\mathrm{sl}}^{k,e}, \pmb{\mathrm{L}}^{k,e}, \pmb{f}^{k,e}}{ \mathrm{min} } \quad \frac{1}{\sum_{b=0}^{B-1} \rs \alpha_b \sum_{u \in \mathcal{U}_b} \mathrm{1}_{u, \mathrm{sl}}^{k,e} \cdot [\alpha_u \mathrm{L}_u^{k,e} ] } \\
    &\mathrm{s.t.} \quad C_1:  \quad \mathrm{1}_{u,\mathrm{sl}}^{k,e} \in \{0,1\}, \quad \forall u, k, e\\
    &\quad C_2: \quad \sum\nolimits_{u \in \mathcal{U}_b} \mathrm{1}_{u,\mathrm{sl}}^{k,e} = Z, \quad \forall b, k, e \\
    &\quad C_3: \quad 1 \leq \mathrm{L}_u^{k,e} \leq \mathrm{L}, ~ \mathrm{L}_u^{k,e} \in \mathbb{Z}^{+}, \quad \forall u, k, e \\
    &\quad C_4: \quad f_u^{k,e} \leq f_{u, \mathrm{max}}, 
    \quad \forall u, k, e\\
    &\quad C_5: \quad \mathrm{1}_{u, \mathrm{sl}}^{k,e} \cdot [ \mathrm{t}_{u,\mathrm{cp}}^{k,e} + \mathrm{t}_{u,\mathrm{up}}^{k,e} ] \leq \mathrm{1}_{u, \mathrm{sl}}^{k,e} \cdot \mathrm{t_{th}}, \quad \forall u, k, e \\
    &\quad C_6: \quad \mathrm{1}_{u, \mathrm{sl}}^{k,e} \cdot [\mathrm{e}_{u,\mathrm{cp}}^{k,e} + \mathrm{e}_{u,\mathrm{up}}^{k,e} ] \leq \mathrm{1}_{u, \mathrm{sl}}^{k,e} \cdot \mathrm{e}_{u, \mathrm{bd}}, \forall u, k, e
\end{align}
\end{subequations}
where $C_1$  and $C_2$ are constraints for the client selection.
Constraints $C_3$ and $C_4$ ensure that the client selects its local iteration and \ac{cpu} frequency within the upper bounds. 
Besides, $C_5$ and $C_6$ are enforced to satisfy the deadline and energy constraints.


Note that we assume the clients share their system information with their associated \ac{bs}.
The \acp{bs} cooperate and solve (\ref{originalOptimProb}) centrally.
Then, each \ac{bs} conveys the optimized parameters to their selected clients.
It is also worth noting that we may equivalently minimize $ - \sum\nolimits_{b=0}^{B-1}\alpha_b \sum\nolimits_{u \in \mathcal{U}_b} \mathrm{1}_{u, \mathrm{sl}}^{k,e} \cdot [\alpha_u \mathrm{L}_u^{k,e} ]$. 
Intuitively, this objective function should select the clients to optimize the weighted combination of their $\mathrm{L}_u^{k,e}$'s. 
However, both (\ref{originalOptimProb}) and this equivalent objective function do not guarantee energy efficiency.
Besides, based on our dataset acquisition and content request model, it is reasonable to consider $\mathrm{t_{th}}$ is at least as long as the duration of the video content. 
Therefore, we seek an energy-efficient solution and consider the following weighted combination of the $\mathrm{L}_{u}^{k,e}$'s and energy expense of the clients as our objective function. 
\begin{align}
\label{utilFunc}
    \varphi_b^{k,e} \coloneqq & - \theta \sum\nolimits_{b=0}^{B-1}\alpha_b \sum\nolimits_{u \in \mathcal{U}_b} \mathrm{1}_{u, \mathrm{sl}}^{k,e} \cdot [\alpha_u \mathrm{L}_u^{k,e} ] + \nonumber\\
    &\qquad (1-\theta)\sum\nolimits_{b=0}^{B-1}\alpha_b \sum\nolimits_{u \in \mathcal{U}_b} \mathrm{1}_{u, \mathrm{sl}}^{k,e} \cdot [\alpha_u \mathrm{e}_{u,\mathrm{tot}}^{k,e} ], 
\end{align}
where $\mathrm{e}_{u, \mathrm{tot}}^{k,e} = \mathrm{e}_{u, \mathrm{cp}}^{k,e} + \mathrm{e}_{u, \mathrm{up}}^{k,e}$ and $\theta \in [0,1]$.

However, the optimization problem is still a binary mixed-integer nonlinear problem and is NP-hard.
Therefore, we first relax the integer constraint on $\mathrm{L}_{u}^{k,e}$, and then define a new variable $\bar{\mathrm{L}}_{u}^{k,e} \coloneqq \mathrm{1}_{u,sl}^{k,e} \cdot \mathrm{L}_{u}^{k,e}$ to replace the multiplication of binary and continuous variables.
\begin{align}
    1 \cdot \mathrm{1}_{u, \mathrm{sl}}^{k,e} &\leq \bar{\mathrm{L}}_{u}^{k,e} \leq \mathrm{L} \cdot \mathrm{1}_{u, \mathrm{sl}}^{k,e}; \quad 0 \leq \bar{\mathrm{L}}_{u}^{k,e} \leq \mathrm{L}. \label{sgdIterC1}\\
    1 \cdot (1 - \mathrm{1}_{u, \mathrm{sl}}^{k,e}) &\leq \mathrm{L}_{u}^{k,e} - \bar{\mathrm{L}}_{u}^{k,e} \leq \mathrm{L} \cdot (1 - \mathrm{1}_{u,\mathrm{sl}}^{k,e}) \label{sgdIterC2}.\\
    \bar{\mathrm{L}}_{u}^{k,e} &\leq \mathrm{L}_u^{k,e} + (1 - \mathrm{1}_u^{k,e}) \mathrm{L} \label{sgdIterC3}.
\end{align}
Besides, we replace the binary client selection variable as  
\begin{align}
    &\sum
    
    _{b=0}^{B-1}\sum_{u \in \mathcal{U}_b} \mathrm{1}_{u,\mathrm{sl}}^{k,e} - \sum_{b=0}^{B-1}\sum_{u \in \mathcal{U}_b} (\mathrm{1}_{u,\mathrm{sl}}^{k,e} )^2 \leq 0; \quad 0 \leq \mathrm{1}_{u,\mathrm{sl}}^{k,e} \le 1.
\end{align}

To that end, we re-write the objective function as 
\begin{align}
    &\tilde{\varphi}_b^{k,e} \coloneqq - \beta \sum_{b=0}^{B-1} \rs \alpha_b \rs \rs \sum_{u \in \mathcal{U}_b} \rs\rs \alpha_u \bar{\mathrm{L}}_u^{k,e} + (1-\beta)\sum_{b=0}^{B-1} \rs \alpha_b \rs \rs \sum_{u \in \mathcal{U}_b} \rs \rs \alpha_u [\mathrm{1}_{u, \mathrm{sl}}^{k,e} \mathrm{e}_{u,\mathrm{up}}^{k,e} + \nonumber\\
    &~ \Tilde{\mathrm{e}}_{u,\mathrm{cp}}^{k,e} ] + \varrho \big( \sum\nolimits_{b=0}^{B-1} \sum\nolimits_{u \in \mathcal{U}_b} \mathrm{1}_{u,\mathrm{sl}}^{k,e} - \sum\nolimits_{b=0}^{B-1}\sum\nolimits_{u \in \mathcal{U}_b} (\mathrm{1}_{u,\mathrm{sl}}^{k,e} )^2 \big), 
\end{align}
where $\Tilde{\mathrm{e}}_{u,\mathrm{cp}}^{k,e} = \bar{\mathrm{L}}_u^{k,e} \times 0.5\zeta \mathrm{n} \bar{\mathrm{n}} \mathrm{c}_u \mathrm{D}_u (f_u^{k,e})^2$ and $\varrho > 0$ is a positive constant that acts as a penalty.
Besides, using first-order Taylor series, we approximate the last quadratic term and rewrite the objective function as
\begin{align}
\label{objFunc}
    &\Bar{\varphi}_b^{k,e} \coloneqq - \beta \sum_{b=0}^{B-1} \alpha_b \rs\rs \sum_{u \in \mathcal{U}_b} \rs \rs \alpha_u \bar{\mathrm{L}}_u^{k,e} + (1-\beta)\sum_{b=0}^{B-1} \rs \alpha_b \rs \rs \sum_{u \in \mathcal{U}_b} \rs \rs \alpha_u [\mathrm{1}_{u, \mathrm{sl}}^{k,e} \mathrm{e}_{u,\mathrm{up}}^{k,e} + \nonumber\\
    &\Bar{\mathrm{e}}_{u,\mathrm{cp}}^{k,e} ] + \varrho \big( \sum_{b=0}^{B-1} \sum_{u \in \mathcal{U}_b} \rs [1 - 2 \cdot \mathrm{1}_{u,\mathrm{sl}}^{k,e,(i)} ] \mathrm{1}_{u,\mathrm{sl}}^{k,e} + \rs \sum_{b=0}^{B-1} \sum_{u \in \mathcal{U}_b} \rs (\mathrm{1}_{u,\mathrm{sl}}^{k,e, (i)} )^2 \big), \rs\rs \rs
\end{align}
where $\bar{\mathrm{e}}_{u,\mathrm{cp}}^{k,e} = \zeta \mathrm{n} \bar{\mathrm{n}} \mathrm{c}_u \mathrm{D}_u f_u^{k,e,(i)} \big[0.5 f_u^{k,e,(i)} \bar{\mathrm{L}}_u^{k,e} + \bar{\mathrm{L}}_u^{k,e,(i)} f_u^{k,e} - f_u^{k,e,(i)} \bar{\mathrm{L}}_u^{k,e,(i)} \big]$. Besides, $\mathrm{1}_{u,\mathrm{sl}}^{k,e,(i)}$, $\bar{\mathrm{L}}_u^{k,e,(i)}$ and $f_u^{k,e,(i)}$ are some initial feasible points. 
Furthermore, we approximate the non-convex computation time as follows
\begin{align}
    \bar{\mathrm{t}}_{u,\mathrm{cp}}^{k,e} 
    \rs \approx [\mathrm{n} \bar{\mathrm{n}} \mathrm{c}_u \mathrm{D}_u/f_u^{k,e,(i)}] \big(\bar{\mathrm{L}}_u^{k,e,(i)} \rs - \bar{\mathrm{L}}_{u}^{k,e,(i)} f_u^{k,e} / f_u^{k,e,(i)} \rs + \bar{\mathrm{L}}_u^{k,e}  \big)\!,\rs\rs 
\end{align}



We thus transform the original optimization problem as
\begin{subequations}
\label{transFormedOptimProb}
\begin{align}
    &\tag{\ref{transFormedOptimProb}}
    \underset{\pmb{\mathrm{1}}_{\mathrm{sl}}^{k,e}, \bar{\pmb{\mathrm{L}}}^{k,e}, \pmb{\mathrm{L}}^{k,e}, \pmb{f}^{k,e}}{ \mathrm{min} } \quad \bar{\varphi}_b^{k,e} \\
    &\mathrm{s.t.} \quad  \quad 0 \leq \mathrm{1}_{u,\mathrm{sl}}^{k,e} \leq 1, \quad \sum\nolimits_{u \in \mathcal{U}_b} \mathrm{1}_{u,\mathrm{sl}}^{k,e} = Z, \quad \forall b, k, e \\
    &\qquad \quad 1 \leq \mathrm{L}_u^{k,e} \leq \mathrm{L}, ~ (\ref{sgdIterC1}), (\ref{sgdIterC2}), (\ref{sgdIterC3}), C_4 \\ 
    &\qquad \quad \bar{\mathrm{t}}_{u,\mathrm{cp}}^{k,e} + \mathrm{1}_u^{k,e} \mathrm{t}_{u,\mathrm{up}}^{k,e} \leq \mathrm{1}_{u, \mathrm{sl}}^{k,e} \cdot \mathrm{t_{th}}, \quad \forall u, k, e \\
    &\qquad \quad \bar{\mathrm{e}}_{u,\mathrm{cp}}^{k,e} + \mathrm{1}_{u, \mathrm{sl}}^{k,e} \cdot \mathrm{e}_{u,\mathrm{up}}^{k,e}  \leq \mathrm{1}_{u, \mathrm{sl}}^{k,e} \cdot \mathrm{e}_{u, \mathrm{bd}}, \forall u, k, e
\end{align}
\end{subequations}
where the constraints are taken for the same reasons as in (\ref{originalOptimProb}).


Note that problem (\ref{transFormedOptimProb}) belongs to the class of ``difference of convex programming" problems and can be solved iteratively using existing tools such as CVX \cite{diamond2016cvxpy}.  
Our proposed iterative solution is summarized in Algorithm \ref{iterAlg}. 
\begin{algorithm}[!t]
\small
\SetAlgoLined 
\DontPrintSemicolon
\KwIn{Initial points $\mathrm{1}_{u,\mathrm{sl}}^{k,e,(i)}$'s, $\bar{\mathrm{L}}_{u}^{k,e,(i)}$, $f_u^{k,e,(i)}$'s, $i=0$, total iteration $I$, precision level $\bar{\varepsilon}$, and $\rho$ \;}
\nl{\textbf{Repeat}: \;} 
\Indp { $i \gets i+1$ \;
        Solve (\ref{transFormedOptimProb}) using $\mathrm{1}_{u,\mathrm{sl}}^{k,e,(i-1)}$'s, $\bar{\mathrm{L}}_{u}^{k,e,(i-1)}$, $f_u^{k,e,(i-1)}$'s and $\rho$, and get optimized $\mathrm{1}_{u,\mathrm{sl}}^{k,e}$'s, $\mathrm{L}_{u}^{k,e}$'s, $\bar{\mathrm{L}}_{u}^{k,e}$'s, $f_u^{k,e}$'s \;
        $\mathrm{1}_{u,\mathrm{sl}}^{k,e,(i)} \gets \mathrm{1}_{u,\mathrm{sl}}^{k,e}$; $\bar{\mathrm{L}}_{u}^{k,e,(i)} \gets \bar{\mathrm{L}}_{u}^{k,e}$, $f_u^{k,e,(i)} \gets f_u^{k,e}$ \;
        }
\Indm \textbf{Until} converge with precision $\bar{\varepsilon}$ or $i=I$\;
\KwOut{Optimal $\mathrm{1}_{u,\mathrm{sl}}^{k,e}$'s, $\mathrm{L}_{u}^{k,e}$'s, $\bar{\mathrm{L}}_{u}^{k,e}$'s and $f_u^{k,e}$'s}
\caption{Iterative Solution for (\ref{transFormedOptimProb})}
\label{iterAlg}
\end{algorithm}


\section{Simulation Results and Discussions}
\subsection{Simulation Setting}
\noindent
To show the effectiveness of the proposed approach, we present simulation results from a system with the following settings. We consider $B=4$ and $U=48$.
The coverage radius of the \ac{bs} is $400$ meters and each \ac{bs} has $|\mathcal{U}_b|=12$ \acp{ue}.
Besides, $G=8$, $\bar{C}=32$, $C=256$, $K=300$, $E=4$, $\mathrm{L}=50$, $\mathrm{n}=10$, $\bar{\mathrm{n}}=32$, $\phi=32$ \cite{pervej2023hierarchical}, $\zeta=2\times10^{-28}$\cite{liu2020client}, $I=50$, $\rho=1$, $\theta=0.4$ and $\kappa=\mathrm{t_{th}}=150$ seconds.
The activity levels $p_{u,\mathrm{ac}}$'s and probability of requesting similar content from the same genre $\upsilon_u$'s are drawn uniformly randomly from $[0.2, 0.8]$ and $[0.1, 0.8]$, respectively.
The genre preferences $p_{u,g}$'s are generated using $\mathrm{Dir}(\pmb{0.3})$ distribution. 
The $c_u$'s, $f_{u,\mathrm{max}}$'s, $\mathrm{e}_{u,\mathrm{bd}}$'s and $\bblue{P_{u,\mathrm{tx}}}$'s are randomly drawn from $[25, 40]$ cycles, $[1.2, 2.0]$ GHz, $[0.8, 1.5]$ Joules and $[20,30]$ dBm, respectively.
Furthermore, $\omega=3 \times 180$ kHz and carrier frequency is $2.4$ GHz.
The path losses and line-of-sight probabilities are modeled based on the urban macro model as listed in \cite[Section $7.4$]{3GPPTR38_901}.
The number of \acp{prb} is varied based on $|\Bar{\mathcal{U}}_b^{k,e}|$'s.
Moreover, we have used a fully connected (FC) neural network\footnote{Our solution is general and can easily be extended to accommodate other neural networks, such as recurrent neural networks or transformers.} that has the following architecture: $\mathrm{FC}(\mathrm{input\_shape}, 512) \rightarrow \mathrm{FC}(512, 256) \rightarrow \mathrm{FC}(256, \mathrm{output\_shape})$. 
Finally, the clients use a sliding window technique, where, in slot $t$, each \ac{ue} processes its dataset so that the feature vector is the previously requested content’s information and the label is the currently requested content’s label.


\begin{figure*}[!t]
\begin{minipage}{0.33\textwidth}
    \centering
    \includegraphics[width=\textwidth]{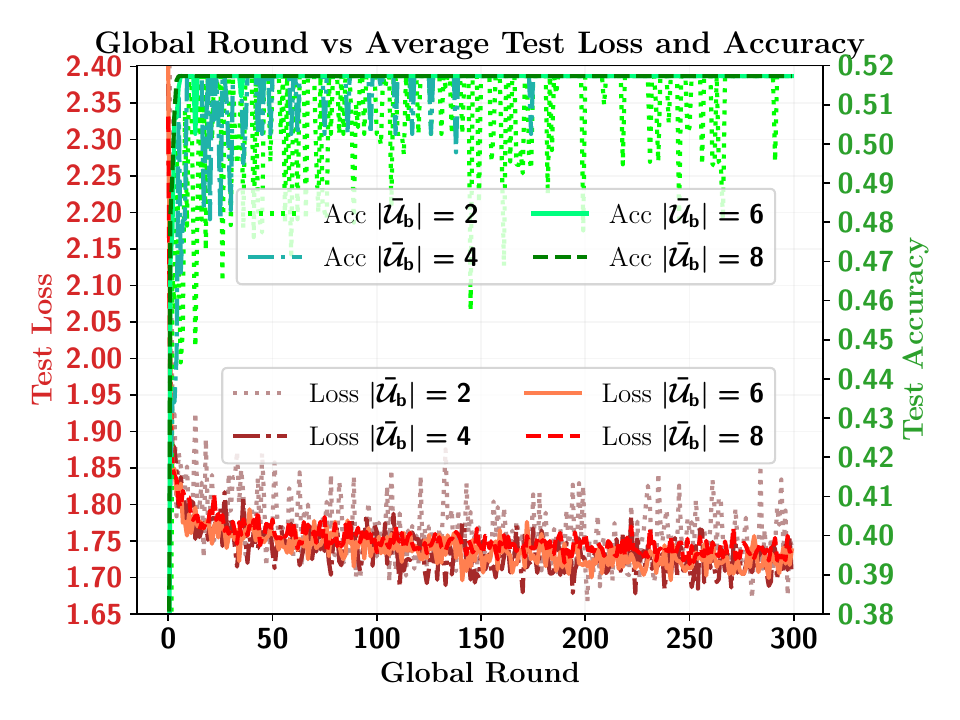}
    \caption{Global round vs average test loss and accuracy}
    \label{flRoundVsTestLossAcc}
\end{minipage} \hspace{0.01in}
\begin{minipage}{0.33\textwidth}
    \centering
    \includegraphics[width=\textwidth]{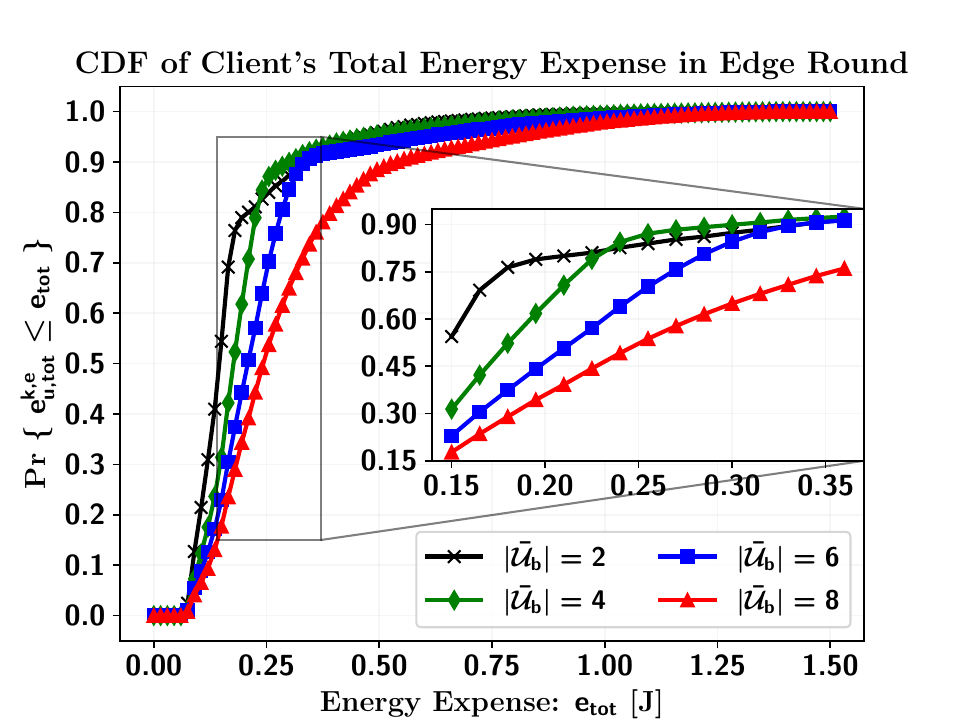}
    \caption{\bblue{CDF of client's total energy expense}}
    \label{cdfEnergyExpense}
\end{minipage} \hspace{0.01in}
\begin{minipage}{0.33\textwidth}
    \centering
    \includegraphics[width=\textwidth]{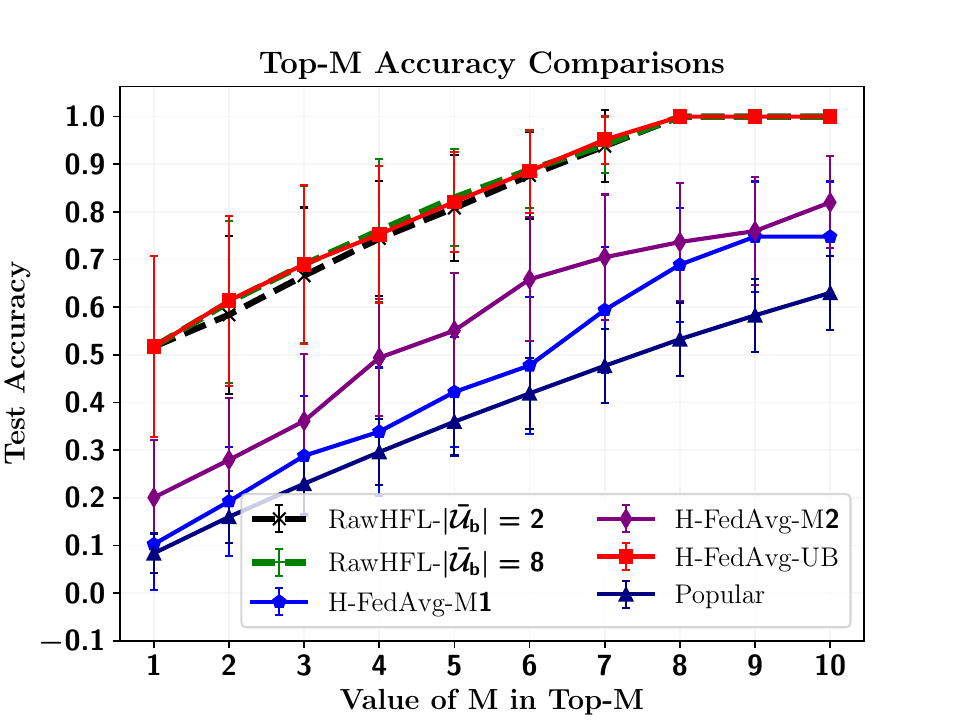}
    \caption{Top-M accuracy comparisons}
    \label{topKacc}
\end{minipage}
\end{figure*}

\subsection{Performance Analysis}
\noindent
We first observe the convergence performance of the proposed algorithm with respect to different $|\bar{\mathcal{U}}_b^{k,e}|$'s in Fig. \ref{flRoundVsTestLossAcc}.
Intuitively, when the selected client set's size is small, the global model is trained on fewer data samples. 
Under severe data heterogeneity, this may lead to poor performance if the clients are not selected appropriately.
Besides, based on the objective function in (\ref{utilFunc}), the proposed solution proactively selects the clients that minimize the weighted utility function.
Therefore, it is expected that \ac{rawhfl} may take more global rounds to convergence when the $|\bar{\mathcal{U}}_b^{k,e}|$ is small.
Our simulation results also show similar trends in Fig. \ref{flRoundVsTestLossAcc}.
We observe that the test loss and test accuracy performances improve when $|\bar{\mathcal{U}}_b^{k,e}|$ increases.
For example, with $|\bar{\mathcal{U}}_b^{k,e}|=2$, the test accuracy drops about $3\%$ when $K$ is around $294$, while the test accuracy reaches a plateau when $K$ is about $175$ with $|\bar{\mathcal{U}}_b^{k,e}|=4$.

However, while a larger $|\bar{\mathcal{U}}_b^{k,e}|$ may help RawHFL convergence faster, the bandwidth and energy costs also increase.
More specifically, when $|\bar{\mathcal{U}}_b^{k,e}|$ increases, (\ref{transFormedOptimProb}) must choose the defined number of clients so that the utility function is minimized. 
As such, it may select some clients with higher energy expenses.
Our simulation results also validate this intuition.
Fig. \ref{cdfEnergyExpense} shows the \ac{cdf} of the total energy expenses during each edge round $e$ for different client set sizes.
For example, the clients spend no more than $0.18$ Joules of energy about \bblue{$76\%$, $52\%$, $37\%$ and $29\%$} of the edge rounds, when $|\bar{\mathcal{U}}_b^{k,e}|=2$, $|\bar{\mathcal{U}}_b^{k,e}|=4$, $|\bar{\mathcal{U}}_b^{k,e}|=6$ and $|\bar{\mathcal{U}}_b^{k,e}|=8$, respectively.

\subsection{Performance Comparisons}
\noindent
We next show performance comparisons with some baselines.
To our best knowledge, no existing baseline exactly  considers our system design and uses \ac{hfl} for video caching.
As such, we modify the traditional \ac{hfedavg} algorithm \cite{liu2020client} for comparison. 
In the modification, termed \ac{hfedavg}-M$1$, we find the smallest number of local rounds that all clients can train their local models without violating their constraints. 
In the second modification, termed \ac{hfedavg}-M$2$, we drop the straggler who cannot even perform a single local round and find the least number of local iterations that the rest of the remaining clients can perform without violating their constraints.
In the third modification, termed \ac{hfedavg}-UB, we consider the upper bound of \ac{hfedavg} \cite{liu2020client}, where each client can perform $\mathrm{L}$ local rounds without constraints. 
We assume $Z=|\mathcal{U}_b|$ for these baselines. 
Finally, we consider a naive popularity-based Top-Popular baseline.

In Fig. \ref{topKacc}, we show the Top-M accuracy comparison of our proposed solution with these baselines.
While a higher value of M should increase the accuracy, the baselines in constrained cases are expected to perform worse. 
Since all clients perform the same number of local rounds in \ac{hfedavg}-M$1$, some \ac{es} may fail to train the model in some edge rounds as some clients may not have sufficient battery powers to offload their trained models due to poor channel conditions.
\ac{hfedavg}-M$2$ should work better than \ac{hfedavg}-M$1$ as we drop these stragglers in \ac{hfedavg}-M$2$.
Besides, \ac{hfedavg}-UB is the ideal case upper bound.
Furthermore, since the clients request content following a popularity-preference tradeoff, the Top-Popular baseline is expected to perform worse.
Our simulation results in Fig. \ref{topKacc}, where the horizontal and the vertical lines show the mean and standard deviation of the test accuracies across all $48$ clients, also validate these trends.
Particularly, our solution yields about $41.5\%$, $31.7\%$ and $43.5\%$ higher (Top-$1$) test accuracies than \ac{hfedavg}-M$1$, \ac{hfedavg}-M$2$ and Top-Popular baselines, respectively.
While \ac{rawhfl} achieves nearly identical test accuracies compared to the \ac{hfedavg}-UB baseline, our solution is energy efficient.
We list the energy expenses of these baselines in Table \ref{performComp}, which clearly indicates that our proposed solution outperforms these baselines.

\begin{table}[!t]
\centering
\caption{Performance comparison: $\!K\!\!=\!400$, $E\!=\!4$, $\mathrm{t_{th}}\!\!=\!150$ s}
\label{performComp}
\fontsize{8}{8}\selectfont
\begin{tabular}{|c|c|c|c|} \hline 
    \textbf{FL Algorithm} & \textbf{Test Accuracy} & \textbf{Energy Expense [J]}  \\ \hline 
    \ac{rawhfl}-$|\Bar{\mathcal{U}}_b|=2$ & $ 0.5172 \pm 0.1894$ & $1757.49$ \\ \hline 
    \ac{rawhfl}-$|\Bar{\mathcal{U}}_b|=4$ & $ 0.5172\pm 0.1894$ & $3961.47$ \\ \hline 
    \ac{rawhfl}-$|\Bar{\mathcal{U}}_b|=6$ & $ 0.5172 \pm 0.1894$ & $7015.58$ \\ \hline 
    \ac{rawhfl}-$|\Bar{\mathcal{U}}_b|=8$ & $0.5172 \pm 0.1894$ & $11587.6$ \\ \hline 
    H-FedAvg-M$1$ & $0.1026 \pm	0.0956$ & $66.53$ \\ \hline 
    H-FedAvg-M$2$ & $0.2006 \pm 0.1205$  & $31275.49$ \\ \hline
    H-FedAvg-UB \cite{liu2020client} & $0.5172 \pm 0.1894$  & $31275.50$ \\ \hline
    Top-Popular & $0.0837 \pm 0.0414$ & N/A  \\ \hline
\end{tabular}
\end{table}

\section{Conclusion}
\noindent
We proposed a privacy-preserving \ac{rawhfl} solution for video caching under a realistic content request model and real-time data sensing mechanism.
Based on our convergence analysis and content request model, we optimized client selections, local training rounds and \ac{cpu} frequencies jointly to minimize a weighted utility function that facilitated faster convergence energy-efficiently.
Moreover, our results suggest a tradeoff between the number of participating clients that facilitates faster convergence and the corresponding resource expenses. 

\medskip
\noindent
\bblue{{\bf Acknowledgments:} The authors thank Dr. Min-Seok Choi, Omer Gokalp Serbetci and Yijing Zhang for the helpful discussions.}  

\bibliography{Reference.bib}
\bibliographystyle{IEEEtran}

\end{document}